\documentclass[aip,jmp,amsmath,amssymb,reprint,]{revtex4-1}

\usepackage[pdftex]{graphicx}
\usepackage{color}

\usepackage{dcolumn}
\usepackage{bm}

\begin{document}

\title{Optical study of lithographically defined, subwavelength plasmonic wires and their coupling to embedded quantum emitters} 

\author{G. Bracher}
\email{gregor.bracher@wsi.tum.de}
\author{K. Schraml}
\author{M. Ossiander}
\author{S. Fr{\'e}d{\'e}rick}
\author{J. J. Finley}
\author{M. Kaniber}

\affiliation{Walter Schottky Institut and Physik Department, Technische Universit\"at M\"unchen, Am Coulombwall 4, 85748 Garching, Germany}

\date{\today}

\begin{abstract}
We present an optical investigation of surface plasmon polaritons propagating along nanoscale Au-wires, lithographically defined on GaAs substrates. A two-axis confocal microscope was used to perform spatially and polarization resolved measurements in order to confirm the guiding of surface plasmon polaritons over lengths ranging from $5-20~\mu m$ along nanowires with a lateral dimension of only $\approx 100~nm$. Finite difference time domain simulations are used to corroborate our experimental observations and highlight the potential to couple proximal quantum emitters to propagating plasmon modes in such extreme sub-wavelength devices. Our findings are of strong relevance for the development of semiconductor based integrated plasmonic and active quantum plasmonic nanosystems that merge quantum emitters with nanoscale plasmonic elements.
\end{abstract}

\maketitle

\section{Introduction}

The fields of active \cite{cao2009active} and quantum \cite{tame2013quantum,chang2006quantum} plasmonics have attracted significant interest  over the past few years \cite{barnes2003} due to the potential to guide and confine light in structures far below the diffraction limit\cite{krenn2002} and probe plasmonic phenomena at the quantum limit. In metal-dielectric nanostructures, tightly localized surface plasmon polaritons (SPP) modes exist that facilitate the confinement of optical signals to nanometer dimensions and propagation velocities close to the speed of light \cite{zia2006plasmonics}. In many experiments presented so far, SPP waveguides were realized using chemically synthesized silver wires with diameters down to a few tens of nanometers \cite{{sanders2006observation},{akimov2007},{ditlbacher2005silver},{dickson2000unidirectional}}. Although, such chemically synthesized nanowires (NWs) show remarkably low losses, deterministic fabrication of integrated structures, such as beam splitters\cite{heeres2009} or interferometers\cite{bozhevolnyi2006channel}, is difficult to achieve. Moreover, deterministic coupling of such elements to proximal quantum emitters is challenging calling for complex post fabrication positioning e.g. of luminescent nano-crystals \cite{huck2011controlled}. In strong contrast, complex and highly integrated plasmonic circuits can be readily fabricated using electron- or ion-beam lithographic techniques. For future integrated quantum optics experiments, it would be highly desirable to combine narrow metallic NWs with nearby optically active quantum emitters\cite{akimov2007}, such as quantum dots\cite{woggon1997optical}, NV-centers\cite{jelezko2006single} or fluorescent molecules\cite{ditlbacher2002a}. In order to observe coupling effects between SPPs and emitters it has been predicted theoretically that NWs with diameters of the order of 100~nm are necessary\cite{chang2006quantum}.
For self-assembled quantum dots (QDs) the optical dipole of the fundamental excitonic transition is commonly orientated parallel to the GaAs surface as a consequence of the heavy hole \cite{heiss2007observation} nature of the valence band state involved that produces TE-polarized light. This orientation is  unfavorable for coupling to SPPs that are purely TM polarized in a two dimensional metal-dielectric interface.
Indeed the coupling of the excitonic dipole to SPPs has been shown to vanish for infinite metallic films on semiconductors \cite{andersen2010strongly}, however,  becoming finite for structures with structured lateral topology where the symmetry is broken.


Here, we demonstrate the fabrication and optical investigations of lithographically defined gold (Au) NWs on GaAs substrates. Using a two-axis confocal microscope we investigated the generation of propagating SPPs from far field radiation by exciting the NW end and image the propagating plasmons by detecting scattered radiation into the far field at the remote nanowire end. Polarization sensitive measurements of SPP propagation with a k-vector orthogonal to the NW axis confirm that plasmons can only be excited from the far field using radiation polarized parallel to the nanowire axis (TM polarization), with a degree of polarization $DoP=\frac{I_{max}-I_{min}}{I_{max}+I_{min}}=73 \pm 2~\%$. The generation, guiding and detection of SPPs in extreme sub-wavelength structures are strongly supported by simulations and in excellent agreement with literature\cite{dickson2000unidirectional} for different material systems.
We also performed numerical simulations to explore the coupling of near surface QDs to the SPPs.  Surprisingly, these simulations show that the structured topology of the metallic nanowires would allow the TE-orientated dipole of near surface QDs to couple to the primarily TM-polarized SPP modes for positions that are highly localized, over sub 50~nm dimensions close to the edges of the nanowire.

\section{Fabrication and Experimental Setup}

\begin{figure}[t!]
\includegraphics[width=0.95\columnwidth]{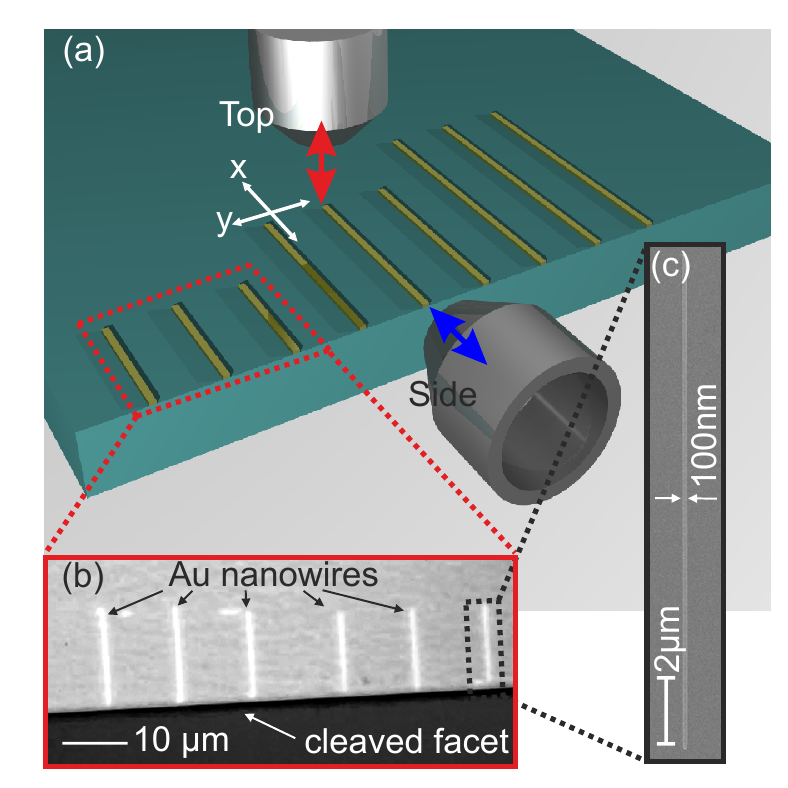}
\caption{(a) Schematic of optical measurement technique: a laser is scanned over the sample. SPPs scattered into the far-field after propagating along the wire are collected by a lateral microscope objective. (b) Microscope image of a cleaved sample: the $100\times100~nm$ wires are fabricated in groups of five wires each having the same dimensions. After fabrication, the sample is cleaved to gain lateral access to the NWs. (c) Scanning electron microscope image of a NW with $L_{NW}=15 \pm 0.2~ \mu m$ and $w_{NW}= 100 \pm 10 ~nm$.}
\label{fig1}
\end{figure}

The sample investigated was defined using electron beam lithography on a nominally undoped (100) GaAs substrate. A bi-layer resist was established by spin-coating a combination of polymethylmethacrylat (PMMA) 200K (AR-P 641.07) and PMMA 950K (AR-P 671.02), that was subsequently baked on a hot-plate at 180$^\circ$C for 6 minutes. This resulted in a total resist thickness of $460\pm10~nm$. After electron beam writing using a Raith e\_LiNE (20~kV, 32~pA, $10~\mu m $ aperture and a dose of $350 \mu As/cm^2$), the sample was developed for 30~s in methylisobutylketone diluted with isopropanol in the ratio 1:3. After this, we deposited a homogeneous layer of 100~nm Au using an electron beam evaporation at a low rate of 0.1~nm/s to ensure a high overall uniformity with an RMS roughness of 2~nm. This layer thickness was chosen in order to suppress leakage radiation into the substrate \cite{drezet2008leakage} and to de-couple the modes at the upper (Au-air) and lower (Au-GaAs) interfaces.  Using this layer thickness the lift-off procedure was found to have a high yield only for wires with a width-to-thickness aspect ratio above unity.  Based on these combined considerations we chose to study NWs with a width of 100~nm and square cross section. Finally, a lift-off was performed to reveal the NWs with a typical width of $w_{NW}= 100 \pm 10 ~nm$ and various lengths $L_{NW}$ ranging from $5~\mu m$ to $20~\mu m$. After structural investigations using atomic force microscopy and scanning electron microscopy (fig. \ref{fig1} (c)), the sample was cleaved perpendicular to the long axis of the NWs. To cleave the sample we score the GaAs surface with a diamond tip under a microscope, enabling a positioning accuracy of the cleaved edge of $\sim 50~\mu m$. Groups of nanowires with different lengths were shifted with respect to each other by $5~\mu m$ giving rise to cleaved and non-cleaved wires near the facet. Although some of the cleaved wire ends were slightly deformed after cleaving, the adhesion was found to be good.

 This provided optical access via the side using a two-axis confocal microscope with full control of the polarization in both the excitation and detection channels. Figure \ref{fig1} (b) shows a microscope image of the cleaved facet: The NWs are separated by $10~\mu m$ and they are arranged in groups of 5 sharing the same length $L_{NW}=5-20~\mu m$ to allow statistics to be obtained.

%

For top excitation, light from a linearly polarized continuous wave Ti:Sapphire laser tuned to 1.49~eV was focused onto the one end of the NW via the top channel using a $50\times$ microscope objective with a numerical aperture $NA = 0.55$, as indicated in figure \ref{fig1} (a). A second $25\times$ microscope objective with a $NA = 0.40$ is placed perpendicular to the cleaved facet, enabling optical access from the side as indicated by the blue arrows in figure \ref{fig1} (a). This geometry enables the sample to be excited from the top \cite{dickson2000unidirectional} while detecting from the side, as well as end-fire excitation of SPPs \cite{stegeman1983excitation} via the side with detection via the top.  

\section{End-fire Excitation}

\begin{figure}[t!]
\includegraphics[width=0.95\columnwidth]{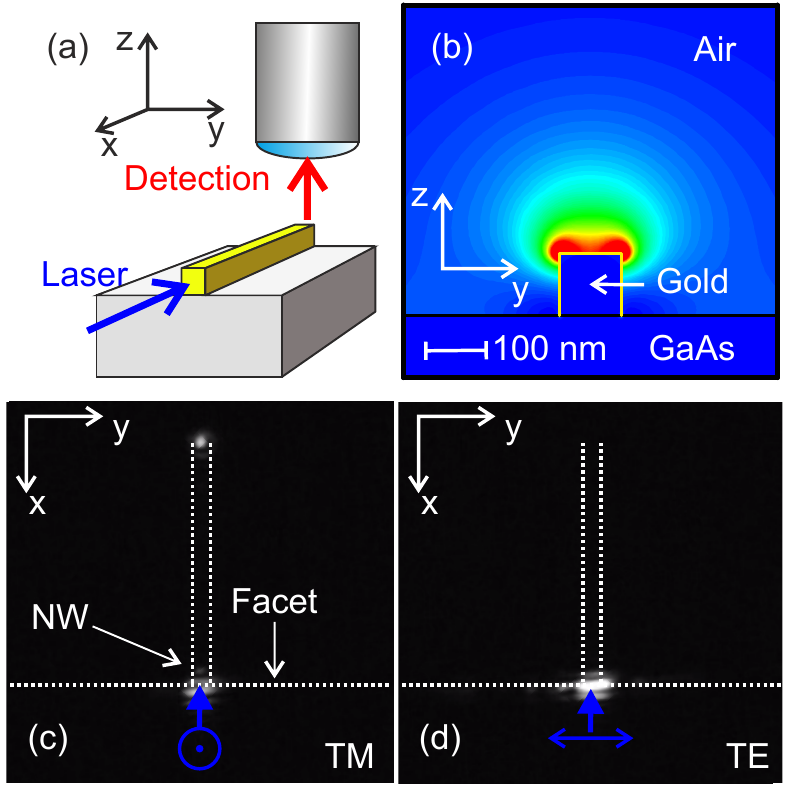}
\caption{(a) Illustration of the end-fire excitation scheme used: SPPs are generated by focussing the laser onto the cleaved facet. At the remote NW end, the SPPs scatter into the far field, where they can be observed via the top channel. (b) Mode simulation of a 100~nm wide waveguide: A highly TM-polarized SPP mode is propagating at the gold-air interface. CCD images recorded in the top channel with TM (c) and TE (d) polarization. Arrows indicate direction of incident laser. A bright spot is observed at the NW end only for TM polarization. }
\label{fig2}
\end{figure}

Figure \ref{fig2} (a) depicts a schematic representation of the end-fire excitation of SPPs, whereby the laser was focused onto the NW at the cleaved facet. The propagating SPPs scatter at the remote NW end and couple to free space optical modes \cite{bracher2011direct}, whereupon they are detected via the top channel. In order to understand the propagation of the SPPs in these NWs, we performed both FDTD \cite{lumfdtd} and mode profile \cite{lummode} calculations. Figure \ref{fig2} (b) depicts the calculated mode at the gold-air interface guided along a $100~nm \times 100~nm$ rectangular NW. Our simulations predict a propagation length of only $2.3~\mu m$, which is probably related to radiation losses into the high refractive index semiconductor substrate. This assumption is supported by similar simulations performed on a glass substrate (n=1.5) that result in a $20 \times$ increased propagation length (data not shown). As expected for all plasmons \cite{novotny2006,Maier2007} the calculated mode is highly TM polarized with a corresponding degree of polarization $DoP = 94~\%$, that would be expected to be manifested in our measurements. Figure \ref{fig2} (c) shows a CCD image recorded via the top channel (red arrow -  figure \ref{fig1} (a)). At the bottom of the CCD image one can clearly see the excitation laser, which is focused onto the cleaved facet. At the remote end of the NW, we observe a bright spot for TM polarized excitation. Due to the k-vector mismatch \cite{barnes2003,Maier2007, novotny2006} in the dispersion relation, the SPPs can couple to the light mode with the help of scatterers and discontinuities \cite{flynn2010,bracher2011direct}. Therefore, the SPPs can be observed as a bright spot only at the NW end. Since we expect a mode with $94~\%$ TM polarization, TE polarized excitation is not expected to result in the generation of propagating SPPs. This expectation is confirmed by figure \ref{fig2} (d) that shows the same NW illuminated with TE polarized light. Both experiment and simulation show that we excite and guide SPPs in our lithographically defined NW using end-fire excitation. 
%

\begin{figure}[t!]
\includegraphics[width=0.95\columnwidth]{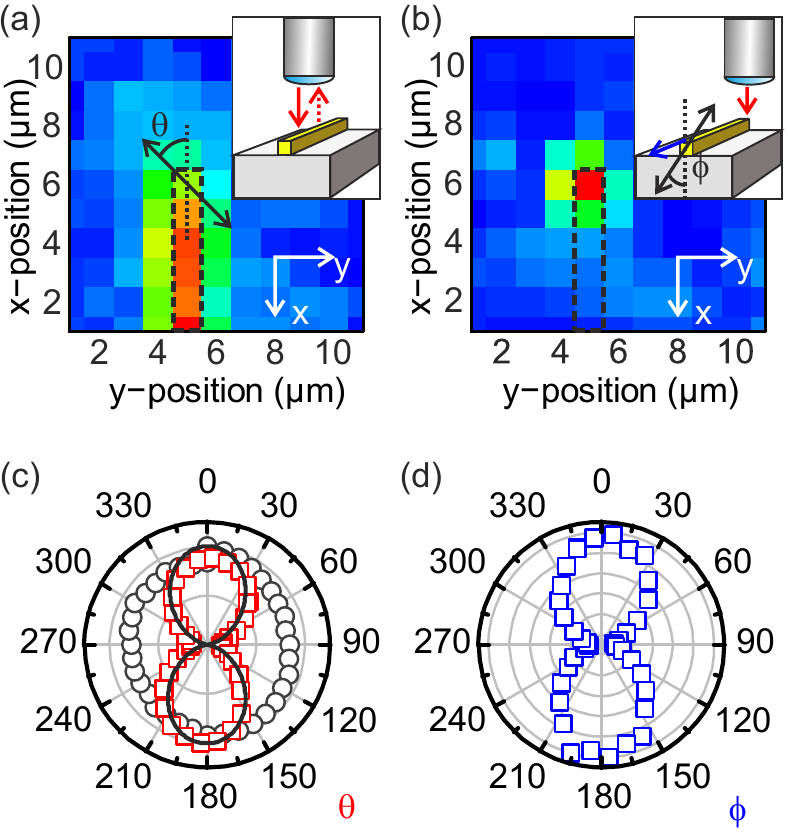}
\caption{(a) Reflectivity map of the NW sample for the determination of the NW position: The end of the $100 \pm 10 ~nm$ wide NW is located at $(x_{top},y_{top})= ( 5.6 \pm 0.2~ \mu m, 4.7 \pm 0.1~\mu m ) $. (b) Output recorded in the side channel is in excellent agreement with the NW end: Highest signal is obtained at $(x_{side}, y_{side})=( 5.9 \pm 0.1~ \mu m, 4.7 \pm 0.1~\mu m )$. (c) Polarization dependent measurement for the excitation: A high degree of polarization $DoP_{top~exc} = 73 \pm 2~\%$ is observed for the NW (red squares), whilst $DoP_{scat}= 10 \pm 2~ \%$ is observed for a dust particle (gray circles). NW behavior is in excellent agreement with simulation (black solid line). (d) Polarization dependent measurement in the detection channel: Collected light is highly TM polarized with $DoP_{side~det} = 95 \pm 2 ~\%$. }
\label{fig3}
\end{figure}

\section{Surface Plasmon Polariton Generation via Scattering}

Since the coupling of the SPPs to the free space optical modes at the NW end is a bi-directional process, we also excited SPPs by focusing the laser onto the NW end along the top channel. Here, we performed spatially resolved reflectivity scans, recording the reflected laser light to locate the NW end and, thus, to identify the position of each NW. Figure \ref{fig3} (a) depicts a typical reflectivity map, where we can clearly identify the NW position, the end positioned at $(x_{top},y_{top})= ( 5.6 \pm 0.2~ \mu m, 4.7 \pm 0.1 ~\mu m ) $. Simultaneously, while scanning the excitation laser over the NW we recorded the output from the side channel. Figure \ref{fig3} (b) shows typical spatially dependent output revealing a single bright spot, only when the excitation spot is close to the NW end. Again we fitted the cross-section to determine the exact position to be at $(x_{side}, y_{side})=(5.9 \pm 0.1 \mu m, 4.7 \pm 0.1~\mu m)$, which is in excellent agreement with the position of the NW end. Taking this observation into account, these findings clearly demonstrate, that we can also couple light to the SPP mode, by focusing the laser via the top onto the NW end. Figure \ref{fig3} (c) shows a polarization dependent measurement for the excitation of a NW (red squares) and a reference measurement of a dielectric scatterer on the sample surface (gray circles). Whilst the scatterer exhibits a very low degree of polarization with $DoP_{scat}= 10 \pm 2~\%$, the NW excitation is highly polarized with $DoP_{top~exc} = 73 \pm 2~\%$. The black solid line in figure \ref{fig3} (c) shows a simulation of the excitation polarization with excellent agreement with experiment. Since only TM polarized light excites SPPs in the end-fire excitation geometry, we also performed a polarization dependent analysis of the light collected in the side channel. Figure \ref{fig3} (d) illustrates that the collected light is highly TM polarized with $DoP_{side~det} = 95 \pm 2 ~\%$, which is in excellent agreement with the mode simulation of figure \ref{fig2} (b). Comparing NWs with different lengths, we obtain a propagation length of $L_{SPP}=3.3 \pm 0.6 ~\mu m$, which is  in fair agreement with our simulations. For a detailed description of the experiment, we refer the reader to the supplementary material.

\begin{figure}[t!]
\includegraphics[width=0.95\columnwidth]{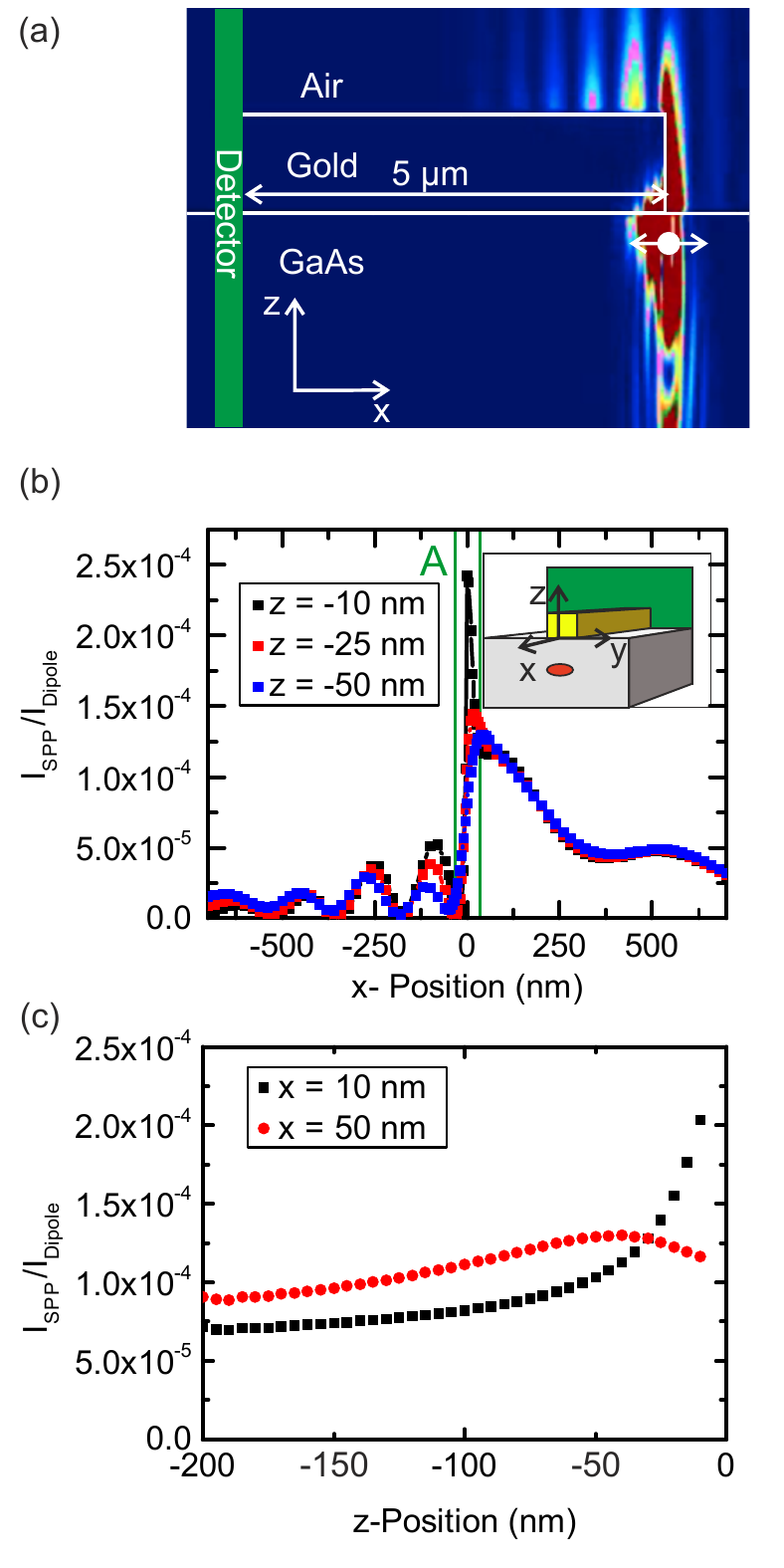}
\caption{(a) FDTD simulation of a dipole emitter located in the GaAs at the end of a $100 \times 100~nm$ NW at $(x=50~nm,y=0~nm,z=-30~nm)$. A bound SPP mode is propagating at the gold-air interface. (b) Plasmon intensity normalized to the dipole intensity as a function of the x-position. Inset: Schematics of the simulation setup. (c) Plasmon intensity normalized to the dipole intensity as a function of the z-position for two different x-positions.}
\label{fig4}
\end{figure}

\section{Coupling between Nanowires and Emitters}

For future quantum plasmonics applications, it is desirable to combine these lithographically defined NWs with proximal emitters such as self-assembled InGaAs quantum dots\cite{finley2001observation,curto2013multipolar}. We continue to demonstrate that highly localized coupling to the propagating SPP mode is possible close to the NW edges. Hereby, we performed finite difference time domain (FDTD)\cite{lumfdtd} simulations to calculate the coupling efficiency between a dipole emitter with a resonant energy $E= 1.3 \pm 0.1~eV$ and a pulse length of $\Delta t = 6~fs$ and a nanometer scale NW. The position of the dipole emitter with respect to the NW end was varied to explore different regimes of coupling. For all simulations, the origin of the Cartesian coordinate system is located at the NW end as schematically depicted in figure \ref{fig4} (b) (inset). We kept the dipole located at the midpoint of the nanowire ($y=0$), since the dipole - SPP coupling is strongest in the center of the WG (data not shown). \linebreak
Figure \ref{fig4} (a) shows a typical FDTD simulation of a $5~\mu m$ long NW with a cross section of $100 \times 100~nm^2$. The emitter is placed at $(x=50~nm,y=0~nm,z=-30~nm)$ with the dipole orientated parallel to the Au-GaAs interface, i.e. simulating a self-assembled quantum dot \cite{andersen2010strongly}. The dipole moment of the quantum emitter is oriented such that it can excite a propagating mode traveling vertically up the gold-air interface at the edge of the NW, along the z-axis. This induces an oscillating dipole at the upper corner of the NW that in turn leads to SPP propagation along the x-direction. The normalized intensity of the SPP is then monitored at the far edge of the NW by measuring the power traversing the $yz-$plane at $x=-5~\mu m$. This normalized intensity is defined as the total power reaching the detector, expressed as a fraction of the total power emitted by the dipole, i.e. $I_{SPP}/ I_{dipole}$. The simulation time was 50~fs, which was sufficient to monitor the entire $6~fs$ pulse in $5~\mu m$ distance. The launching of the SPP along the vertical surface is the reason, why the dipole is able to excite SPPs in our structures, despite the disadvantageous orientation of the dipole moment \cite{andersen2010strongly}. Furthermore, to the quantitative simulations of the sole generation of SPPs by near-surface emitter, we performed spatially resolved simulations to understand the detailed coupling mechanisms of the emitter. Figure \ref{fig4} (b) shows $I_{SPP}/I_{dipole}$ coupled into the SPP mode propagating along the x-direction as a function of the x-position of the dipole. The various curves presented in figure \ref{fig4} (b) were calculated for $z=-10~nm$ (black symbols), $z=-25~nm$ (red symbols) and $z=-50~nm$ (blue symbols). When the dipole is located directly underneath the NW ($x<0$) we observe intensity oscillations as x decreases. These oscillations result from an interference of SPPs propagating in negative x-direction with SPPs being launched in the positive x-direction and reflected at the NW end. The lack of a edge discontinuity to which the TE-orientated dipole can couple leads to a very low in-coupling efficiency ($<5.0 \times 10^{-5} \cdot I_{SPP}$), which decreases further with an  increasing distance to the NW end and is only weakly dependent on the z-distance. Although the intensity reduces from  $5.0 \times 10^{-5} \cdot I_{SPP}$ to $2.0 \times 10^{-5} \cdot I_{SPP}$ as the z-position of the dipole is increased from $z=-10~nm$ to $z=-50~nm$ at $x=-100~nm$, all three curves converge for $x<-300~nm$. For such large displacements the absolute distance to the NW end does not depend strongly on the z-position. Therefore, we conclude that such a convergence is a clear indication for far-field coupling, the radiated power from the dipole exciting the SPP. For $z= -10~nm$ and $x>0$ we observe a pronounced peak close to $x=10~nm$, labeled A on figure \ref{fig4} (b). This indicates near-field SPP generation, where $5 \times$ more power is coupled into the waveguide  ($I_{SPP}=2.5\times 10 ^{-4} \cdot I_{dipole}$) and detected at the NW end. Considering the expected losses of such a $5~\mu m$ long NW with a propagation length of $2.3~\mu m$, the out of plane position \cite{chang2006quantum} and the wrong dipole orientation \cite{andersen2010strongly}, we estimate that a surprisingly large fraction of the overall dipole intensity, nearly  $0.2\%$, couples to the SPP mode. As the separation of the dipole from the metal-semiconductor interface increases further to $z=-25~nm$ we find that this pronounced peak reduces from $2.5 \times 10^{-4}$ to $1.5 \times 10^{-4}$ and is slightly shifted by $\Delta x = +15~nm$. Increasing the vertical distance further to $z=-50$ this peak vanishes completely. For larger displacements in the x-direction ($x>100~nm$) we find again a periodic oscillation for all z-distances. This oscillation is probably related again to far-field excitation of SPPs and results from the spatial displacement to the NW end. In contrast, the amplitude of the highly resonant peak A at $z=-10~nm$ and $x=10~nm$ decreases very strongly as the distance of the dipole from the NW end increases. This strong localization relates to the strong SPP confinement at the metal-dielectric interface \cite{Maier2007,barnes2003} and therefore, to near-field coupling. Figure \ref{fig4} (c) shows the detected SPP power as a function of the z-dependence, for two different lateral displacements ($x= 10~nm$ and $x= 50~nm$). Again, the SPP intensity slowly varies in the z-direction for $x=50~nm$. However, for $x=10~nm$ we find strong increase in the SPP intensity from $1.1 \times 10^{-4}$  to $2.5 \times 10^{-4}$  within $\Delta z =50~nm$. The simulations presented in figure \ref{fig4} (c) are indicative of two different coupling regimes between the dipole and the SPP mode; for $z>-50~nm$ with a strong dependence of the coupling efficiency on both x and z position, and for $z<-50~nm$ where the coupling efficiency is only weakly dependent on z. The strong localization of the effective SPP generation can be explained by coupling of the emitter to the NW via the near-field. This type of highly local coupling of a dipole emitter with the SPP mode holds much promise for on-chip quantum optics experiments\cite{akimov2007,heeres2009} with the potential to excite nano-structure via SPPs over nanometer length scales.

\section{Conclusion}

In summary, we have presented all optical investigations of SPPs  propagation along lithographically defined, sub wavelength plasmonic nanowires on GaAs. End-fire excitation of the SPP modes on the NW was shown to be possible only for TM polarized light, which is in excellent agreement with mode simulations, where we calculated the propagation of a $94\%$ TM polarized mode with a propagation length of $2.3~\mu m$ along a $100 \times 100~ nm^2$ rectangular NW. Both spatially resolved and polarization dependent measurements showed that the excitation and detection path can be freely interchanged as expected. For both the excitation and the detection we measured high degrees of polarization with $DoP_{top~exc} = 73 \pm 2~\%$ and $DoP_{side~det} = 95 \pm 2 ~\%$, respectively. Finally, we performed simulations of an emitter in the vicinity of the NW end, which can couple to the SPP mode with $0.2~\%$ efficiency. Detailed studies on the positional dependence of the coupling efficiency gave clear indications for near-field and highly local excitation of SPPs providing much promise for deterministic quantum plasmonic experiments.

\begin{acknowledgments}
We acknowledge financial support of the DFG via the SFB 631, Teilprojekt B3, the German Excellence Initiative via NIM, and FP-7 of the European Union via SOLID.
G. Bracher gratefully acknowledges the support of the TUM Graduate School's Faculty Graduate Center Physik at the Technische Universit\"at M\"unchen.
\end{acknowledgments}

%

\end{document}